# Crystallography of deformation twinning in hexagonal close-packed metals. Revisiting the case of the (56°, **a**) contraction twins in magnesium

Cyril Cayron

Laboratory of ThermoMechanical Metallurgy (LMTM), PX Group Chair, Ecole Polytechnique Fédérale de Lausanne (EPFL), Rue de la Maladière 71b, 2000 Neuchâtel, Switzerland

## Abstract

Contraction twinning in magnesium alloys leads to new grains that are misoriented from the parent grain by a rotation (56°, **a**). The classical theory of deformation twinning doesn't precise the atomic displacements and does not explain why contractions twinning is less frequent than extension twinning. The paper proposes a new model in the continuity of our previous works on martensitic transformations and extension twinning. A continuous angular distortion matrix that transforms the initial hcp crystal into a final hcp crystal is determined such that the atoms move as hard spheres and reach the final positions expected by the orientation relationship. The calculations prove that the distortion is not a simple shear when it is considered in its continuity. The $(0\bar{1}11)$ is untilted and restored, but it is not fully invariant because some interatomic distances in this plane evolve during the distortion process; the unit volume also increases up to 5% before coming back to its initial value when the twinning distortion is complete. Then, the distortion takes the form a simple shear on the $(0\bar{1}11)$ plane with a shear direction along the direction $[18\,\bar{5}\,\bar{5}]_{hex}$ and a shear amplitude $\gamma \approx 0.358$. It is the first time that this twinning mode is reported. Experiments are proposed to validate or infirm the new model.

## 1. Introduction

The $\{10\bar{1}1\}$ contraction twins in magnesium and other hexagonal close-packed metals and alloys are formed when the parent grains are compressed along their c-axis. They are less frequent than extension twins and even scarcer than the so-called $\{10\bar{1}1\}\{10\bar{1}2\}$ double-twins, but they can play an important role on the fracture properties. They are observed in transmission electron microscopy (TEM) as thin plates for which diffraction patterns obtained in twin/parent areas exhibit a mirror symmetry through a $\{10\bar{1}1\}$ plane, as it was reported in a bulk polycrystalline Mg-Al-Zn alloy (AZ31) [1], and in a single crystal pure magnesium with bulk [2] or in nano-sized [3] samples. The $\{10\bar{1}1\}$ contraction twins were also evidenced in the Electron Back Scatter Diffraction (EBSD) orientation maps on in AZ31 magnesium alloys by drawing the special grain boundaries between the twins and the surrounding matrix [4][5][6]. The misorientation between the parent grains and their twins is a rotation of 56° around the **a**-axis, simply noted here (56°,**a**).

The $\{10\bar{1}1\}$ contraction twins are usually considered as a shear along a $\{10\bar{1}1\}$ plane in the $\langle 10\bar{1}\bar{2}\rangle$ direction with a shear amplitude given by the formula [7][8]

$$s = \frac{4\left(\frac{c}{a}\right)^2 - 9}{4\sqrt{3}\frac{c}{a}} \qquad (1)$$

For an ideal ratio $c/a = 2\sqrt{6}/3 \approx 1.633$ ratio corresponding a hard-sphere stacking, $s = \frac{5\sqrt{2}}{48} \approx 0.147$. It is not easy to trace back the history of formula (1). Christian and Mahajan [7] cites Reed-Hill's work published in 1960 [9], but only a low shear value is mentioned in paper [9], and it seems actually that this formulae and its associated scheme should be attributed to H.S. Rosenbaum in his chapter "*Nonbasal Slip in h.c.p. Metals and its Relation to Mechanical Twinning*" published in a book edited by Reed-Hill [10]. One can wonder why this twinning mode, despite its very low shear amplitude, is not in the list of the twins calculated for titanium by Crocker and Bevis [11] from a general theory based on shear and rational lattice correspondences [12][13]. The difference of c/a ratio ($c/a \approx 1.588$ for titanium) does not seem to be the explanation. Thus, in order to get a first answer, let us consider Fig. 1 and Fig. 2 that represent a 3D hexagonal unit cell and its projection along the **a**-axis, respectively. More precisely, Fig. 2 shows a supercell 1x1x2 constituted of two unit cells doubled along the **c**-axis in order to make appear the $(0\bar{1}11)$ plane as the diagonal of 1x1x2 supercell. The orientation of the twinned lattice is deduced from the parent lattice by mirror symmetry though this plane. The classical theory of twinning searches for the shear that restores the lattice and put it in mirror symmetry of the parent lattice. Clearly, it predicts the shear vector **s**$_1$ along the $-[122]_{hex}$ direction (this direction written in four indices is indeed of type $\langle 10\bar{1}\bar{2}\rangle_{hex}$), and a shear amplitude ratio $\gamma_1 = \frac{s_1}{h} = \frac{23\sqrt{2}}{24} \approx 1.355$, as illustrated in Fig. 2b. The details of calculation are given in Annex 1. Instead of using the vector **s**$_1$, Rosenbaum used the vector **s**$_2$, as illustrated in Fig. 2b. The calculations are reported in Annex 1. Indeed, the shear amplitude associated to **s**$_2$ is $\gamma_2 = \frac{s_2}{2h} = \frac{5\sqrt{2}}{48} \approx 0.147$, which is exactly the value deduced from formula (1) with $c/a = 2\sqrt{6}/3$. Therefore, we conclude that formula (1) was based on a scheme similar to that of Fig. 2b, even if no trace of this scheme could be found in literature by the present author. Unfortunately the vector **s**$_2$ is not a conventional twinning shear vector because it doesn't restore the 1x1x2 lattice (the rectangle just becomes a distorted rectangle but not a new one), as already noticed by Guo et al. [14]. Rosenbaum used actually a larger supercell which is four times bigger than the 1x1x2 cell (more explanations will be given in discussion). It is true that the lattice of this big supercell is restored by a shear of vector **s**$_2$, but it is absolutely not clear how the atoms move (shuffle) inside this supercell during the lattice distortion. Actually, Yoo [15] mentioned that 7/8 of the atoms move by shuffling in the $\{10\bar{1}1\}\langle 10\bar{1}\bar{2}\rangle$ twinning mode, but their trajectories are absolutely unknown. The reader can convince himself of the difficulty of guessing these trajectories from Fig. 2c. It seems legitimate to question the validity of a model that can only predict 1/8 of the displacements of the atoms. The unconventionally large supercell and associated high shuffling component of the twinning mode probably explains why this mode was not predicted by the initial Crocker and Bevis theory. Indeed, the criteria "*the shuffling mechanism should be simple, the shuffle magnitude should be small, and shuffles should be parallel to the twinning direction rather than perpendicular to this direction*" mentioned in Ref. [12] are not fulfilled since they are totally unknown in the model based on Fig. 2b and related equation (1). This first analysis of literature shows that there is a disagreement for the $\{10\bar{1}1\}$ contraction twins between the general classical theory of twinning and the Rosenbaum's scheme. In addition, neither of them can explicitly describe how the magnesium atoms move in the twinning. There is also another puzzling point. As previously mentioned, experimental observations have shown that the $\{10\bar{1}1\}$ contraction twins are far less frequent and their domains thinner than the extension $\{10\bar{1}2\}$ twins. Why the morphologies are so different if both modes have close shear values (0.147 and 0.118, respectively) ?

From all these points, it is clear that the crystallography of $\{10\bar{1}1\}$ contraction twins in Mg needs important clarifications, or even more, a new basis. In a previous paper [16], it was shown that extension twins in hexagonal close-packed metals such as magnesium deviate from simple shear by steric effect due to the atom diameters (taken as hard-spheres in a first approximation). It was shown that a volume change of 3% occurs during the lattice distortion and that even the twinning plane can't be let fully invariant during the distortion process (it could just be let untilted and restored at the end of the distortion). A energy criterion was introduced to generalize the Schmid law for non-shear matrices; it predicts the formation of extension for some orientations where the parent crystal is orientated such that the Schmid factor is negative; which could explain the apparent "anomalous" character of extension twinning sometimes reported in literature. The aim of the present paper is to follow the same approach to revisit the crystallography of $\{10\bar{1}1\}$ contraction twins. The calculations will lead to a shear matrix for the complete distortion that is different from the Rosenbaum's one and that was not predicted by the classical theory of twinning. In addition, the volume change associated to the distortion will be calculated and compared to the extension twinning one. An orientation graph that predicts the formation of the contraction twins will be given. It will show very large differences between the Rosenbaum's model and the new proposed model.

## 2. Notations and calculation rules

In first approximation the atoms are assumed to be hard-spheres of constant diameter. The ratio of lattice parameters taken in the calculations is the ideal hcp ratio:

$$c/a = 2\sqrt{6}/3 \qquad (2)$$

The three-index notation in the hexagonal system is preferentially chosen for the calculations. The planes will be sometimes written in four-index notation, but mainly to refer to literature. The vectors are noted by bold lowercase letters and the matrices by bold capital letters.

We call $\mathbf{B}_{hex} = (\mathbf{a}, \mathbf{b}, \mathbf{c})$ the usual hexagonal basis, and $\mathbf{B}_{ortho} = (\mathbf{x}, \mathbf{y}, \mathbf{z})$ the orthonormal basis represented in Fig. 1 and linked to $\mathbf{B}_{hex}$ by the coordinate transformation matrix $\mathbf{H}_{hex}$:

$$\mathbf{H}_{hex} = [\mathbf{B}_{ortho} \to \mathbf{B}_{hex}] = \begin{pmatrix} 1 & -1/2 & 0 \\ 0 & \sqrt{3}/2 & 0 \\ 0 & 0 & c/a \end{pmatrix} \qquad (3)$$

In order to follow the displacements of the Mg atoms during extension twinning, some labels are given to the atomic positions, as illustrated in Fig. 1. The labels are the same as those of ref. [16]. Some nodes will not be used here (such as U, V, S, T), and other nodes were added. We note O, the "zero" position that will be let invariant by the distortion. We call X, Y and Z the atomic positions defined by the vectors **OX** = ***a*** = [100]$_{hex}$, **OY** = ***a*** + 2***b*** = [120]$_{hex}$ and **OZ** = ***c*** = [001]$_{hex}$. It can be checked by using the matrix **H**$_{hex}$ that **OX** = [100]$_{ortho}$, **OY** = [0 $\sqrt{3}$ 0]$_{ortho}$ and **OZ** = [0 0 c/a]$_{ortho}$. The atom in the center of the face (O, X, Y, S) is noted M$_1$, and the atom close to the face (O, X, Z, T) is noted N; their vectors are **OM**$_1$ = [110]$_{hex}$ and **ON** = [$\frac{2}{3}, \frac{1}{3}, \frac{1}{2}$]$_{hex}$. At the same z-level as N, there are the atoms P and Q given by **OP** = [$\frac{5}{3}, \frac{4}{3}, \frac{1}{2}$]$_{hex}$ and **OQ** = [$\frac{2}{3}, \frac{4}{3}, \frac{1}{2}$]$_{hex}$. The other atoms that will be used in the paper are given by the vectors **OM**$_2$ = [010]$_{hex}$ = [$\frac{-1}{2}, \frac{\sqrt{3}}{2}, 0$]$_{ortho}$, **OF** = [0, $\frac{-\sqrt{3}}{3}$, 0]$_{ortho}$, **OW** = [$\frac{1}{2}, \frac{\sqrt{3}}{2}$, c/a]$_{ortho}$ and **OE** = [$\frac{-1}{2}, \frac{-\sqrt{3}}{2}$, c/a]$_{ortho}$.

The model is the same as for extension twins [16] but the atomic displacements were more difficult to determine and the calculations were more complex. In order to automatize them, a mathematical procedure was developed. It consists in determining the trajectory of an atom X where X = [x,y,z]$_{ortho}$, such that X keeps contact with three atoms A, B, C, and the trajectories of these three atoms are known as analytical functions of a unique parameter β. The atoms A, B, C, and X have the same diameter equal to 1. The problem consists in finding the solution(s) to the equations

$$\begin{cases} \|X-A\| = 1 \\ \|X-B\| = 1 \\ \|X-C\| = 1 \end{cases}, i.e \begin{cases} (x-x_A(\beta))^2 + (y-y_A(\beta))^2 + (z-z_A(\beta))^2 = 1 \\ (x-x_B(\beta))^2 + (y-y_B(\beta))^2 + (z-z_B(\beta))^2 = 1 \\ (x-x_C(\beta))^2 + (y-y_C(\beta))^2 + (z-z_C(\beta))^2 = 1 \end{cases} \qquad (4)$$

Manual calculations are always possible but very heavy for some complex trajectories. Sometimes, geometrical considerations can help finding more quickly the solutions. For example, the point X should be on the circle separating the segment AB, i.e. a circle that has for centre the middle of the segment [AB], the vector **AB** for normal, and of radius $\sqrt{1-\frac{AB^2}{4}}$, which equals $\frac{\sqrt{3}}{2}$ if A and B keep contact during their displacements. Most of the time, the calculations remain heavy. That is why equation (4) was systematically solved in its symbolic form by using the solver of Mathematica. For sake of simplicity, this procedure will be called X = FourthAtom (A,B,C).

In some cases, the displacements of some atoms X will be determined by imposing a contact only with two atoms A and B. Thus, only the two first equations of list (4) can be used. The third equation then comes from another condition, generally by imposing that the displacement of the atom X occurs in a specific plane, for example x = 0 if the displacement occurs in the OYZ plane. This procedure will be noted X = FourthAtom (A,B, x=0).

## 3. New orientation obtained by an atomic rotation in the basal plane

In first lecture, the reader only interested in the lattice distortions (and not in the mechanism and atomic displacements) can go directly to section 5.

Before calculating the distortion related to the case in which the atoms of the twinned crystal are exactly the mirror positions through the $(0\bar{1}11)$ plane, i.e the exact (56°,**a**) twinning mode, we propose to consider another configuration slightly misoriented from the exact one. The main idea comes from the observation that the atom Q is close to the image of Z through the $(0\bar{1}11)$ mirror plane, the distance OQ is close to the distance OZ, and that **OQ** is normal to **OF**, as illustrated in Fig. 3. We have therefore imagined, after many unfruitful attempts, a series of atomic displacements that seem reasonable because they are small, they respect the hard-sphere conditions, and they restore the hcp lattice at the end of the process. We can't prove that this model gives the best solution, but the solution that emerges does not suffer from the crippling problems mentioned in the introduction. The reasoning follows a sequence of displacements:

- The atoms O and X do not move. **OX** gives the invariant line of the distortion.
- The atom $M_1$ in the $(001)_{hex}$ plane and initially in the **a** + 2**b** position, moves such that the vector **OM₁** rotates by 30° to become perpendicular to the **OX** line when the transformation is complete. It remains on the $(001)_{hex}$ plane during its rotation. The angle β = (**OX**, **OM₁**) increases continuously from 60° to 90°.
- The atom $M_2$ follows the same trajectory as $M_1$ translated by the vector –**OX**, i.e. **OM₂** = **OM₁**–**OX**.
- By steric effect, the movement of the atom $M_1$ displaces the atom N such that it keeps contact with atoms O, X and $M_1$, i.e. its trajectory is given by the solution of N = FourthAtom (O,X,$M_1$).
- The displacement of the atom N displaces the atom F such that F moves relatively to O as N moves relatively to $M_1$, i.e. **OF** = **M₁N** = **ON**–**OM₁**.
- The displacement of the atoms $M_1$ and N makes move the atom Q. We assume that the atom Q keeps contact with the atoms $M_1$ and N and that the displacement occurs in the plane x = 0, i.e. its trajectory is given by the solution of Q = FourthAtom ($M_1$,N,x=0).
- We suppose that the atom Y moves such that it keeps contact with the atoms $M_1$, $M_2$ and Q, i.e. its trajectory is given by the solution of Y = FourthAtom ($M_1$,$M_2$,Q).
- The displacement of the atom F makes move the atom Z such that Z keeps contact with the atoms O and F while remaining on the initial OYZ plane (x=0); i.e. its trajectory is given by the solution of Z = FourthAtom (O,F, x=0).
- The atom E follows the same trajectory related to O as the atom Z related to $M_1$, i.e. **OE** = **OZ**–**OM₁**.

The angle β between the invariant line **OX** and the direction **OM₁** changes continuously from $β_s$ = 60° (start) to $β_f$ = 90° (finish). We note $\kappa = Cos(\beta)$. The parameter κ changes continuously from $κ_s$ = ½ (start) to $κ_f$ =0 (finish). According to the sequence of displacements described previously, the trajectories of all the atoms described can be expressed as a function of $κ$. The calculations performed in the orthonormal basis $\mathbf{B}_{ortho} = (\mathbf{x}, \mathbf{y}, \mathbf{z})$ are not detailed for sake of simplicity. They lead to:

$$\mathbf{OX}(\kappa) = [1, 0, 0] \tag{5}$$

$$\mathbf{OM_1}(\kappa) = [\kappa, \sqrt{1-\kappa^2}, 0] \tag{6}$$

$$\mathbf{OM_2}(\kappa) = [\kappa - 1, \sqrt{1-\kappa^2}, 0] \tag{7}$$

$$\mathbf{ON}(\kappa) = \left[\frac{1}{2}, \quad \frac{1}{2}\sqrt{\frac{1-\kappa}{1+\kappa}}, \quad \frac{1}{\sqrt{2}}\sqrt{\frac{1+2\kappa}{1+\kappa}}\right] \tag{8}$$

$$\mathbf{OF}(\kappa) = \left[\frac{1}{2} - \kappa, \quad -\frac{(1+2\kappa)}{2}\sqrt{\frac{1-\kappa}{1+\kappa}}, \quad \frac{1}{\sqrt{2}}\sqrt{\frac{1+2\kappa}{1+\kappa}}\right] \tag{9}$$

$$\mathbf{OQ}(\kappa) = \left[0, \quad \frac{4}{(3-2\kappa)}\sqrt{\frac{1-\kappa}{1+\kappa}}, \quad \frac{2\sqrt{2}(1-\kappa)}{(3-2\kappa)}\sqrt{\frac{1+2\kappa}{1+\kappa}}\right] \tag{10}$$

$$\mathbf{OY}(\kappa) = \left[\kappa - \frac{1}{2}, \quad \frac{11 + 4(1-\kappa)\kappa}{2(3-2\kappa)}\sqrt{\frac{1-\kappa}{1+\kappa}}, \quad \frac{1-2\kappa}{\sqrt{2}(3-2\kappa)}\sqrt{\frac{1+2\kappa}{1+\kappa}}\right] \tag{11}$$

$$\mathbf{OZ}(\kappa) = \left[0, \quad 0, \quad \frac{\sqrt{2(1+\kappa)(1+2\kappa)}}{1+\kappa}\right] \tag{12}$$

$$\mathbf{OE}(\kappa) = \left[-\kappa, \quad -\sqrt{1-\kappa^2}, \quad \frac{\sqrt{2(1+\kappa)(1+2\kappa)}}{1+\kappa}\right] \tag{13}$$

The column vectors are written here in line for editorial reason. They are all expressed in the basis $\mathbf{B}_{ortho}$.

For extension twinning the distortion matrix was calculated with the unit cell XYZ; however, here, this supercell is not appropriate because the calculations with this cell show that its volume at the end of the distortion is not the same as it was at the beginning. For contraction twinning, the unit cell that conserves the volume is XYE. The three column vectors of this cell, **OX**, **OY**(κ) and **OE**(κ) form a basis $\mathbf{B}_{XYE}$ given by the coordinate transformation matrix:

$$[\mathbf{B}_{ortho} \rightarrow \mathbf{B}_{XYE}(\kappa)] = \mathbf{B}_{XYE}(\kappa) \tag{14}$$

$$= \begin{pmatrix} 1 & \kappa - \frac{1}{2} & -\kappa \\ 0 & \frac{11 + 4(1-\kappa)\kappa}{2(3-2\kappa)}\sqrt{\frac{1-\kappa}{1+\kappa}} & -\sqrt{1-\kappa^2} \\ 0 & \frac{1-2\kappa}{\sqrt{2}(3-2\kappa)}\sqrt{\frac{1+2\kappa}{1+\kappa}} & \frac{\sqrt{2(1+\kappa)(1+2\kappa)}}{1+\kappa} \end{pmatrix}$$

The continuous distortion matrix at each step κ of the process is given in the basis $\mathbf{B}_{ortho}$ by the matrix $\mathbf{F}_{ortho}^{p \rightarrow t}(\kappa) = \mathbf{B}_{XYE}(\kappa) \cdot \mathbf{B}_{XYE}^{-1}(\kappa_s)$ (see equation 1 of Ref. [17]). The calculation leads to

$$\mathbf{F}_{ortho}^{p \to t}(\kappa) = \begin{pmatrix} 1 & \dfrac{2\kappa - 1}{2\sqrt{3}} & \dfrac{1}{8}\sqrt{\dfrac{3}{2}}(1 - 2\kappa) \\ 0 & \dfrac{11 + 4(1-\kappa)\kappa}{2\sqrt{3}(3-2\kappa)}\sqrt{\dfrac{1-\kappa}{1+\kappa}} & -\dfrac{\sqrt{3}}{8}\dfrac{(1-4\kappa^2)}{(3-2\kappa)}\sqrt{\dfrac{1-\kappa}{1+\kappa}} \\ 0 & \dfrac{1-2\kappa}{\sqrt{6}(3-2\kappa)}\sqrt{\dfrac{1+2\kappa}{1+\kappa}} & \dfrac{\sqrt{3}}{8}\left(\dfrac{13-10\kappa}{3-2\kappa}\right)\sqrt{\dfrac{1+2\kappa}{1+\kappa}} \end{pmatrix} \quad (15)$$

The ratio of volume change $\mathcal{V}'/\mathcal{V}$ between the unit volume $\mathcal{V}$ in the initial state and its value $\mathcal{V}'$ at any step of the distortion is directly given by the determinant of the distortion matrix:

$$\frac{\mathcal{V}'}{\mathcal{V}}(\kappa) = det(\mathbf{F}_{ortho}^{p \to t}) = \frac{3\sqrt{(1+2\kappa)(1-\kappa)}\,(4 + \kappa(1-2\kappa))}{4(1+\kappa)(3-2\kappa)} \quad (16)$$

The curve of the volume change is reported in Fig. 4. The maximum value is obtained for the intermediate parameter $\kappa_i = 1/4$, i.e. exactly at the midpath of distortion process, and the corresponding volume ratio is then $\frac{\mathcal{V}'}{\mathcal{V}}(\kappa_i) = \frac{297}{200\sqrt{2}} \approx 1.050$, which means that the unit volume increases by 5% during the distortion before coming back to its initial value. Part of this volume change comes from the evolution of the distance OW. Indeed, the distance $\|\mathbf{OW}(\kappa)\| = \|\mathbf{F}_{ortho}^{p \to t}(\kappa).\mathbf{OW}\|$ calculated from the matrix (15) is indeed not constant. The ratio of length OW' by its initial value OW is given by

$$\frac{OW'}{OW}(\kappa) = \sqrt{\frac{3}{41}}\sqrt{\frac{41 + 35\kappa - 70\kappa^2}{3 + \kappa - 2\kappa^2}} \quad (17)$$

The graph is given in Fig. 4. The maximum value is also obtained for the mid-path state $\kappa_i = 1/4$, and the corresponding distance ratio is $\frac{OW'}{OW}(\kappa_i) = \frac{33}{5\sqrt{41}} \approx 1.031$. This means that the distance OW increases by 3% during the distortion before coming back to its initial value.

## 4. Compensation of the obliquity angle

The configuration described previously leads to an hcp structure with an orientation very close to the one that is the mirror image through the $(0\bar{1}11)$ plane. A slight tilt angle exists between these two configurations. Indeed, the direction $\boldsymbol{d} = \left[0, \frac{\sqrt{3}}{2}, \frac{c}{a}\right]_{ortho} = \left[\frac{1}{2}, 1, 1\right]_{hex}$ that lies on the $(0\bar{1}11)$ plane (i.e projection of the direction **OW** on the OYZ plane) slightly rotates during the distortion. Let us call $\xi = (\boldsymbol{d}, \boldsymbol{d}')$, the angle between the initial position of the vector $\boldsymbol{d}$ and its image by the distortion matrix (15). This angle is determined by $(\boldsymbol{d}_{ortho}, \mathbf{F}_{ortho}^{p \to t} \boldsymbol{d}_{ortho})$. The calculations show that its cosine $C_\xi$, is a function of κ:

$$C_\xi(\kappa) = \frac{56 + 15\sqrt{\dfrac{1-\kappa}{1+2\kappa}} - 6\kappa\left(8 - \sqrt{\dfrac{1-\kappa}{1+2\kappa}}\right)}{\sqrt{41}(3 - 2\kappa)} \sqrt{\frac{3 + 4(1-\kappa)\kappa}{41 + 35(1-2\kappa)\kappa}} \quad (18)$$

As shown in Fig. 5, the angle $\xi$ increases from 0 to 1.5° before decreasing down to 1.1°.

The rotation matrix that compensates the angle $\xi$ all along the distortion process is given in the basis $\mathbf{B}_{ortho}$ by

$$\mathbf{R}(\kappa) = \begin{pmatrix} 1 & 0 & 0 \\ 0 & C_\xi(\kappa) & \sqrt{1 - C_\xi(\kappa)^2} \\ 0 & -\sqrt{1 - C_\xi(\kappa)^2} & C_\xi(\kappa) \end{pmatrix} \tag{19}$$

The continuous distortion matrix compensated at each step $\kappa$ of the transformation is given in the basis $\mathbf{B}_{ortho}$ by the matrix $\mathbf{D}_{ortho}^{p \to t}(\kappa) = \mathbf{R}(\kappa) . \mathbf{F}_{ortho}^{p \to t}(\kappa)$, with $\mathbf{R}(\kappa)$ given by equation (19) and $\mathbf{F}_{ortho}^{p \to t}(\kappa)$ given by equation (15). The atomic displacements inside the XYE cell (shuffles) can also be calculated by applying the rotation matrix (19) to the trajectories described by equations (5)-(13).

As the analytical expressions of each of the 3x3 terms of the matrix $\mathbf{D}_{ortho}^{p \to t}(\kappa)$ as function of $\kappa$ are longer than a page width and are not reported in the paper, but they can be found in the Mathematica programs added in the Supplementary Materials. Two important states are however important and their numerical values worth being written. They are the intermediate state obtained at the maximum volume change, i.e. for $\kappa_i = 1/4$, and the final state obtained for $\kappa_f = 0$. The corresponding matrices are

$$\mathbf{D}_{ortho}^{p \to t}(\kappa_i) = \begin{pmatrix} 1 & -\dfrac{1}{4\sqrt{3}} & \dfrac{\sqrt{\dfrac{3}{2}}}{16} \\ 0 & \dfrac{3(11 + 24\sqrt{2})}{20\sqrt{41}} & \dfrac{27(11 - 8\sqrt{2})}{80\sqrt{82}} \\ 0 & \dfrac{-27 + 22\sqrt{2}}{10\sqrt{41}} & \dfrac{9(44 + 9\sqrt{2})}{80\sqrt{41}} \end{pmatrix} \approx \begin{pmatrix} 1. & -0.144 & 0.077 \\ 0. & 1.053 & -0.012 \\ 0. & 0.064 & 0.997 \end{pmatrix} \tag{20}$$

$$\mathbf{D}_{ortho}^{p \to t} = \mathbf{D}_{ortho}^{p \to t}(\kappa_f) = \begin{pmatrix} 1 & -\dfrac{1}{2\sqrt{3}} & \dfrac{\sqrt{\dfrac{3}{2}}}{8} \\ 0 & \dfrac{87}{82} & -\dfrac{15}{328\sqrt{2}} \\ 0 & \dfrac{10\sqrt{2}}{123} & \dfrac{77}{82} \end{pmatrix} \approx \begin{pmatrix} 1. & -0.289 & 0.153 \\ 0. & 1.061 & -0.032 \\ 0. & 0.115 & 0.939 \end{pmatrix} \tag{21}$$

These matrices can be expressed in the hexagonal basis $\mathbf{B}_{hex}$ by using the formula of coordinate change

$$\mathbf{D}_{hex}^{p \to t}(\kappa) = \mathbf{H}_{hex}^{-1} \mathbf{D}_{ortho}^{p \to t}(\kappa) \mathbf{H}_{hex} \tag{22}$$

with $\mathbf{H}_{hex}$ in equation (3). They become:

$$\mathbf{D}_{hex}^{p\to t}(\kappa_i) = \begin{pmatrix} 1 & -\dfrac{5}{8} + \dfrac{9\sqrt{\dfrac{2}{41}}}{5} + \dfrac{33}{40\sqrt{41}} & \dfrac{205 + 99\sqrt{41} - 72\sqrt{82}}{1640} \\ 0 & \dfrac{3(11 + 24\sqrt{2})}{20\sqrt{41}} & \dfrac{9(11 - 8\sqrt{2})}{20\sqrt{41}} \\ 0 & \dfrac{3(-27 + 22\sqrt{2})}{40\sqrt{82}} & \dfrac{9(44 + 9\sqrt{2})}{80\sqrt{41}} \end{pmatrix}$$

$$\approx \begin{pmatrix} 1. & -0.099 & 0.114 \\ 0. & 1.053 & -0.022 \\ 0. & 0.034 & 0.997 \end{pmatrix}$$

(23)

$$\mathbf{D}_{hex}^{p\to t} = \mathbf{D}_{hex}^{p\to t}(\kappa_f) = \begin{pmatrix} 1 & -\dfrac{9}{41} & \dfrac{9}{41} \\ 0 & \dfrac{87}{82} & -\dfrac{5}{82} \\ 0 & \dfrac{5}{82} & \dfrac{77}{82} \end{pmatrix} \approx \begin{pmatrix} 1. & -0.22 & 0.22 \\ 0. & 1.061 & -0.061 \\ 0. & 0.061 & 0.939 \end{pmatrix}$$

(24)

## 5. Analysis of the matrices of complete distortion

The fact that the matrix of complete transformation calculated from complex analytical expression with irrational values gets eventually rational values when expressed in the hexagonal basis is a satisfying result. In addition, the matrix (24) is a shear matrix. It has only 1 as eigenvalue and its eigenspace is of dimension 2, formed by the vectors $[100]_{hex}$ and $[011]_{hex}$, i.e. it is the expected $(0\bar{1}11)_{hex}$ plane. The shear vector and amplitude can be calculated by working in $\mathbf{B}_{ortho}$ and applying the matrix (21) to the vector normal to the $(0\bar{1}11)_{hex}$ plane, $\mathbf{n} = \left[0, -\dfrac{c}{a}, \dfrac{\sqrt{3}}{2}\right]_{ortho}$:

$$\mathbf{s} = (\mathbf{D}_{ortho}^{p\to t} - \mathbf{I}).\mathbf{n} = \dfrac{1}{\sqrt{6}}\left[\dfrac{41\sqrt{3}}{48}, \dfrac{-5}{16}, \dfrac{-5\sqrt{2}}{12}\right]_{ortho} = \dfrac{1}{24}[18\ \bar{5}\ \bar{5}]_{hex}$$

(25)

and the shear amplitude is

$$\gamma = \dfrac{\|\mathbf{s}\|}{\|\mathbf{n}\|} = \dfrac{1}{12}\sqrt{\dfrac{37}{2}} \approx 0.358$$

(26)

To our knowledge, this is the first time that such a twinning mode is proposed. This mode has all the characteristics of the $(0\bar{1}11)$ contraction twinning, i.e. a shear matrix with a $\{10\bar{1}1\}$ plane for invariant plane, reasonable shear amplitude, and more interestingly, all the atomic shuffles are clearly defined by equations (5) to (13) compensated by the rotation (19). The shear matrix (24) was deduced from the atomic displacements and not from shear lattice correspondence as in the classical theory. As in our previous work on extension twinning [16], we point out again the mathematical fact that, although the distortion matrix of complete transformation is a shear matrix, the exact mechanism of deformation twinning can't be a shear if the hard-sphere assumption is adopted. Indeed, we have shown in Fig. 4 that during the distortion process, the volume evolves and that the vector **OW** that belongs to the $(0\bar{1}11)$ plane is not fully invariant because of its length change. The unit volume comes back to its initial value and the $(0\bar{1}11)$ plane is restored when the transformation is complete, but the path taken to reach this restored state can't be ignored. As the $\{10\bar{1}2\}$ extension twins, the $\{10\bar{1}1\}$ contraction twins do not result from a simple shear distortion.

Two other interesting matrices can be deduced from the model: the correspondence matrix, which gives the images by the distortion of the basis vectors of the parent crystal expressed in the twin basis, and the coordinate transformation matrix, which gives the coordinates of the basis vectors of the twin crystal expressed in the parent basis. In the present case, the correspondence matrix can be calculated by considering the vectors used to build the XYE cell (Fig. 3a)

OX: $\mathbf{a}_p \to \mathbf{a}'_p = -\mathbf{a}_t$ (27)

OY: $\mathbf{a}_p + 2\mathbf{b}_p \to (\mathbf{a}_p + 2\mathbf{b}_p)' = -\frac{1}{2}(\mathbf{a}_t + 2\mathbf{b}_t) + \mathbf{c}_t$

OE: $-(\mathbf{a}_p + \mathbf{b}_p) + \mathbf{c}_p \to -(\mathbf{a}_p + \mathbf{b}_p)' + \mathbf{c}'_p = \mathbf{a}_t + 2\mathbf{b}_t$

This means that the basis $\mathbf{B}_{XYE}$ formed by the three vectors $\mathbf{a}_p$, $\mathbf{a}_p + 2\mathbf{b}_p$ and $-(\mathbf{a}_p + \mathbf{b}_p) + \mathbf{c}_p$, written in column, is transformed by distortion into a matrix formed by three new vectors (marked by the prime) that are the vectors of the twinned crystal $-\mathbf{a}_t$, $-\frac{1}{2}(\mathbf{a}_t + 2\mathbf{b}_t) + \mathbf{c}_t$ and $\mathbf{a}_t + 2\mathbf{b}_t$ forming a new basis $\mathbf{B}'_{XYE}$. The negative sign of $-\mathbf{a}_t$ is added to impose that both $\mathbf{B}_{XYE}$ and $\mathbf{B}_{(XYE)'}$ have a positive determinant. The basis $\mathbf{B}_{XYE}$ expressed in the basis $\mathbf{B}^p_{hex}$ and the basis $\mathbf{B}_{(XYE)'}$ expressed in the basis $\mathbf{B}^t_{hex}$ are respectively given by

$$\mathbf{B}^p_{XYE} = \begin{pmatrix} 1 & 1 & -1 \\ 0 & 2 & -1 \\ 0 & 0 & 1 \end{pmatrix} \quad \text{and} \quad \mathbf{B}^t_{(XYE)'} = \begin{pmatrix} -1 & -1/2 & 1 \\ 0 & -1 & 2 \\ 0 & 1 & 0 \end{pmatrix}$$ (28)

Let us call $\mathbf{T}^{p \to t}_{hex}$ the coordinate transformation matrix from the parent to the twin crystal. The basis $\mathbf{B}^t_{(XYE)'}$ expressed in the basis $\mathbf{B}^p_{hex}$ is $\mathbf{T}^{p \to t}_{hex} \mathbf{B}^t_{(XYE)'}$. The two matrices (28) can thus be expressed in $\mathbf{B}^p_{hex}$ and are linked by the equation

$$\mathbf{D}^{p \to t}_{hex} \mathbf{B}^p_{XYE} = \mathbf{T}^{p \to t}_{hex} \mathbf{B}^t_{(XYE)'}$$ (29)

which can be written equivalently

$$\mathbf{D}^{p \to t}_{hex} \left(\mathbf{C}^{t \to p}_{hex}\right)^{-1} = \mathbf{T}^{p \to t}_{hex}$$ (30)

$$\text{with } \mathbf{C}^{t \to p}_{hex} = \mathbf{B}^t_{(XYE)'} \left(\mathbf{B}^p_{XYE}\right)^{-1}$$ (31)

Formula (30) is actually a general formula already introduced in Ref. [17]. The calculations lead to

$$\mathbf{C}_{hex}^{t \to p} = \begin{pmatrix} -1 & \frac{1}{4} & \frac{1}{4} \\ 0 & -\frac{1}{2} & \frac{3}{2} \\ 0 & \frac{1}{2} & \frac{1}{2} \end{pmatrix} \text{ and } \mathbf{T}_{hex}^{p \to t} = \begin{pmatrix} -1 & \frac{9}{41} & \frac{23}{82} \\ 0 & -\frac{23}{41} & \frac{64}{41} \\ 0 & \frac{18}{41} & \frac{23}{41} \end{pmatrix} \qquad (32)$$

The matrix $\mathbf{C}_{hex}^{t \to p}$ is the correspondence matrix. It gives in the twin reference basis the images of the vectors of the parent crystal initially written the parent reference basis. More explicitly, any vector **u** expressed in $\mathbf{B}_{hex}^{p}$, written $\mathbf{u}_{/\mathbf{B}_{hex}^p}$, has for image by the distortion a vector **u'** whose coordinates in the twin basis are given by $\mathbf{u'}_{/\mathbf{B}_{hex}^t} = \mathbf{C}_{hex}^{t \to p} \mathbf{u}_{/\mathbf{B}_{hex}^p}$. It can be checked for example that any vector **u** that belongs to the $(0\bar{1}11)$ plane of the parent crystal, i.e. of type $[x\, y\, y]_{hex}$ when written in $\mathbf{B}_{hex}^{p}$, is transformed by distortion into a new vector **u'** of type $\left[-x + \frac{y}{2}, y, y\right]_{hex}$ when written in $\mathbf{B}_{hex}^{t}$, i.e. that belongs to the $(0\bar{1}11)$ plane of the twin crystal.

The matrix $\mathbf{T}_{hex}^{p \to t}$ is the matrix of coordinate transformation between the parent and twin bases. The calculations show that it is a rotation matrix of 180° around the $[\frac{1}{4}, 1, 1]_{hex}$ axis, which indeed belongs the $(0\bar{1}11)$ plane. The matrices equivalent to $\mathbf{T}_{hex}^{p \to t}$ are obtained by multiplying $\mathbf{T}_{hex}^{p \to t}$ by the matrices of internal symmetry **g** of the hexagonal phase, i.e. the matrices forming the point group of the hcp phase $\mathbb{G}^{hcp}$ (more details can be found in Ref. [18]).

$$\langle \mathbf{T}_{hex}^{p \to t} \rangle = \{ \mathbf{T}_{hex}^{p \to t} \mathbf{g}, \ \mathbf{g} \in \mathbb{G}^{hcp} \} \qquad (33)$$

One interesting matrix is obtained with

$$\mathbf{g} = \begin{pmatrix} -1 & 1 & 0 \\ 0 & 1 & 0 \\ 0 & 0 & -1 \end{pmatrix} \text{; it is } \mathbf{T}_{hex}^{p \to t} = \begin{pmatrix} 1 & -\frac{32}{41} & \frac{105}{82} \\ 0 & -\frac{23}{41} & \frac{64}{41} \\ 0 & -\frac{18}{41} & -\frac{23}{41} \end{pmatrix} \qquad (34)$$

This a rotation matrix of rotation axis $\mathbf{a}_p$ and of angle $\text{ArcCos}\left[-\frac{23}{41}\right] \approx 124.12°$, which has for commentary angle the expected value of 55.88° ≈ 56°. This validates the internal coherency of our calculations.

## 6. Discussion

The Rosenbaum's model detailed in the introduction assumes that the contraction (56°, **a**) twins in magnesium result from a shear along the direction $-[122]_{hex}$ on the $(0\bar{1}11)$ plane with an amplitude $\gamma_2 \approx 0.147$. The supercell used for the calculations is (**OX**, 2 **OZ** − **OY**, 2 **OZ** + **OY**) of volume $8\sqrt{2}$, which is eight time the volume of Bravais cell (**a**, **b**, **c**). Two questions remains unanswered in this model: how move the atoms in such a big super cell and why contraction twins are far less frequent than extension twins if their shear amplitude is relatively close to the extension one? The present paper revisits the subject of contraction (56°, **a**) twinning by using another supercell, i.e. the XYE cell of volume of $2\sqrt{2}$, as for the orthorhombic cell XYZ previously used for extension twinning [16]. The atoms at the corner of the cell (each of them count for 1/8 in the cell) follow the continuous distortion $\mathbf{D}_{ortho}^{p \to t}(\kappa) = \mathbf{R}(\kappa).\mathbf{F}_{ortho}^{p \to t}(\kappa)$, with $\mathbf{F}_{ortho}^{p \to t}(\kappa)$ and $\mathbf{R}(\kappa)$ given in equations (15) and (19), and with the angular parameter that varies from the start value $\beta_s$ = 60° ($\kappa_s$ = ½) to the finish value $\beta_f$ = 90° ($\kappa_f$ =0). The atoms M, N and Q follow the shuffling equations (6), (8) and (10) compensated by the rotation (19), and they count for ½, 1 and ½, respectively. This means that contraction twinning is obtained with 1/3 of distortion and 2/3 of shuffle, as for extension twinning [16]. A 3D movie showing the atomic displacements of a crystal made of 4x4x4 XYE cells is reported in the Supplementary Materials. Three snapshots taken from the initial, intermediate and final states are extracted in Fig. 6. When the distortion is complete, the distortion matrix takes the shape of a simple shear matrix on the $(0\bar{1}11)$ plane, but the shear direction $[18\,\bar{5}\,\bar{5}]_{hex}$ and amplitude $\gamma \approx 0.358$ are given here for the first time. The $(56°, \mathbf{a})$ contraction twinning leads to an exchange a) between the prismatic and $(0\bar{1}13)$ planes (see how the vector **OZ** and **OQ** change in Fig. 3a), and b) between the basal and $(01\bar{1}1)$ planes (see how the vector **OY** and **OE** change in Fig. 3a), while the $(0\bar{1}11)$ plane maintains untilted. In comparison, the extension twins occur such that the prismatic and basal planes are "exchanged" while maintaining the $(0\bar{1}12)$ plane untilted. The fact that the volume change and the shear amplitude of contraction twins are significantly larger that of extension twins (nearly +70% and +200%, respectively) explains why the former are less frequently observed than the latter.

As already introduced for extension twinning [16], an energy criterion that generalizes the Schmid criterion can be used to predict the twin formation with non-shear matrices. This criterion assumes that an energy barrier must be overcome at the maximum volume change, i.e. at the intermediate state at $\kappa = \kappa_i$. The interaction work W of a unit volume of a material that deforms by mechanical twinning inside an external stress field $\mathbf{\Gamma}$ is given by the inner product

$$W = \mathbf{\Gamma}_{ij} . \mathcal{E}_{ij} \tag{35}$$

The work is performed by the external stress during the transformation. A high value of interaction work means a high probability of transformation, and negative value should correspond to an impossibility of transformation. The interaction work is proportional to the Schmid factor in the case of a simple shear [16]. A Mg parent crystal is tilted by an angle $\phi$ around the **x**-axis, and rotated by an angle $\theta$ around the direction **n** normal to $(0\bar{1}11)$ plane. The matrix of rotation of angle $\phi$ around the **x**-axis, and the matrix of rotation of angle $\theta$ around the **n**-axis are noted $\mathbf{R}_x(\phi)$ and $\mathbf{R}_n(\theta)$, respectively (see Ref. [16] for the details). A distortion matrix $\mathbf{D}^{p \to t}$ of the tilted-rotated parent crystal, expressed in the basis $\mathbf{B}_{ortho}$, becomes

$$\mathbf{D}^{p \to t}(\phi, \theta) = \mathbf{R}_x(\phi)\,\mathbf{R}_n(\theta).\mathbf{D}^{p \to t}.\left(\mathbf{R}_x(\phi)\,\mathbf{R}_n(\theta)\right)^{-1} \tag{36}$$

The associated twinning deformation matrix is $\mathcal{E}^{p \to t}(\phi, \theta) = \mathbf{D}^{p \to t}(\phi, \theta) - \mathbf{I}$. The interaction work with a stress field $\mathbf{\Gamma}$ becomes

$$W\,(\phi, \theta) = \mathbf{\Gamma}_{ij}\,.\mathcal{E}_{ij}^{p \to t}(\phi, \theta) \tag{37}$$

Let us do the calculations for different matrices $\mathbf{D}^{p \to t}$: a) the usual Rosenbaum's shear matrix **ShearR** given in Appendix A, and b) the new matrices we have found, i.e. the intermediate and complete distortion matrices $\mathbf{D}_{ortho}^{p \to t}(\kappa_i)$ in equation (20) and $\mathbf{D}_{ortho}^{p \to t}(\kappa_f)$ in equation (21), respectively. The graph of interaction works calculated for the Rosenbaum's matrix and for the intermediate and final states, $W_R(\phi, \theta)$, $W_i(\phi, \theta)$ and $W_f(\phi, \theta)$ are given in Fig. 7 in the case of a uniaxial compression along the **z**-axis of -100 MPa. Contrarily to extension twinning [16], there is no important differences between the works calculated with the intermediate and the final distortion matrices (compare Fig. 7a with Fig. 7b). However, very large difference in the positions of the maxima and minima are found between the works calculated with Rosenbaum's shear matrix **ShearR** (Fig. 7c) and the one calculated with the final distortion matrix (Fig. 7a). A maximum is positioned in the ɸ-axis at $\phi = -0.298 \text{ rad} = -17°$, which comes from the fact that, whatever the model, the $(0\bar{1}11)$ shear plane is oriented at 62° far from the basal plane, i.e. a rotation of -17° is required to place it at 45° from the compression axis. However, the maximum is positioned in the θ-axis at $\theta = 0$ rad for the Rosenbaum's shear matrix and at $\theta = -1.15 \text{ rad} = -66°$ for the final distortion matrix. This difference can be explained by the fact that the shear vector in the Rosenbaum's model is along $-[122]_{hex}$, i.e. it is in the OYZ plane, which is the reference plane ($\theta = 0$) for the calculations of the work, whereas the shear vector found in the present model is along $[18\,\bar{5}\,\bar{5}]_{hex}$, which is at 66° far from the $-[122]_{hex}$ one. Therefore, the two models differ by 66° in their predictions of the formation of contractions twinning. It thus seems possible to confirm or infirm the new model by experimental compression tests on single crystals orientated such that a $\{10\bar{1}1\}$ plane is at 45° of the compression axis and by rotating the crystal around the normal to the chosen $\{10\bar{1}1\}$ plane. A difference of 66° in the prediction is higher than all the uncertainties relative to such type of experiments.

The approach we have followed for the last years on martensitic transformation and more recently on deformation twinning is simple; it consists in calculating how a crystallographic structure can be restored knowing the final orientation and assuming that the atoms moves as hard-spheres. There is no fitting parameter, which makes the model rigid by also very robust. Despite its simplicity and rigidity, the theory appears very fruitful. The fact that a new twinning mode emerges from the calculations in the case of the contraction twinning encourages us in following this path. It will be always possible in the future to refine the model by using molecular dynamics, by calculating with finite element methods the stress field created in the surrounding matrix, by considering the interface dislocations or disconnections resulting from the distortion, etc. More results are obtained on new twinning modes and on the so-called $\{10\bar{1}1\}\{10\bar{1}2\}$ double-twins; they will be the subject of other publications.

## 7. Conclusion

The crystallography of (56°, **a**) contraction twinning in magnesium is revisited. The classical model uses a large supercell with a volume that is eight time that of the Bravais cell. The atomic displacements of the atoms are unknown and the scarcity of the contraction twins in comparison with the extension twins remains unexplained. The present model solves these issues. It is based on a general approach already used for martensitic transformations and extension twinning. It consists in finding a continuous angular distortion that transforms the initial crystal into the final crystal such that the atoms move hard spheres. Generally, an orientation very close the expected orientation relationship is obtained. The exact orientation is then obtained by introducing an additional rotation to compensate the "obliquity" angle. In the case of contraction twinning, this angle is at maximum 1.5°. The calculations are based on a supercell that is only twice that of the Bravais cell. The analytical equations of all the atomic displacements are determined. It is proved that when the distortion is considered in its continuity, it is not a simple shear. The $(0\bar{1}11)$ plane is untilted and restored but it is not fully invariant during the distortion process because some interatomic distances are not constant. In addition, the volume increases up to 5% before coming back to its initial value when the twin is formed. This volume change is higher than that of extension twinning (3%). The final distortion takes the form a shear matrix that has never been reported in the past. It is a simple shear on the $(0\bar{1}11)$ plane with a shear direction along $[18\,\bar{5}\,\bar{5}]_{hex}$ and a shear amplitude $\gamma \approx 0.358$. A generalized Schmid criterion predicts the formation of contraction twins in single crystal for orientations in which these twins should not form according to the classical Rosenbaum's model. Therefore, it seems possible to experimentally validate or infirm the new theoretical model of contraction twinning.

# Note

The Mathematica programs used for the calculation of the distortion matrices, correspondence and coordinate change matrices and for the generalized Schmid factors are given in Supplementary Materials.

# Acknowledgments

I would like to show my gratitude to Prof. Roland Logé, director of LMTM, and to PX group for the laboratory subsidy and for our scientific and technical exchanges.

# Appendix A :

# Calculation of the shear amplitudes in the classical shear theory

Let consider Fig. 2. The diagonal length is the hypotenuse of the rectangle triangle of sides $\sqrt{3}$ and $2c = \frac{8}{\sqrt{6}}$, it is $d = \sqrt{\frac{41}{3}}$. The angle α in the triangle is such that $(\alpha) = \frac{8}{\sqrt{82}}$. The distance h is $h = Sin(\alpha)\sqrt{3}$. We also have $L = \frac{d}{2Cos(\alpha)} = \frac{41}{6\sqrt{3}}$. The intercept theorem proves that $\frac{s_1}{d} = \frac{L-\sqrt{3}}{L}$.

The associated shear amplitude $\gamma_1$ is given by the ratio $\gamma_1 = \frac{s_1}{h} = \frac{23\sqrt{2}}{24} \approx 1.355$. The intercept theorem also proves that $\frac{s_2}{d} = \frac{L-2\sqrt{3}}{L}$. The associated shear amplitude $\gamma_2$ is given by the ratio $\gamma_2 = \frac{s_2}{2h} = \frac{5\sqrt{2}}{48} \approx 0.147$.

The shear vector $\mathbf{s_2}$ is orientated in the direction $-\left[0, \frac{\sqrt{3}}{2}, \frac{2\sqrt{6}}{3}\right]_{ortho}$ such that its norm is equal to $s_2 = d\frac{L-2\sqrt{3}}{L} = \frac{5}{\sqrt{123}} \approx 0.451$. Thus, $\mathbf{s_2} = -\left[0, \frac{5\sqrt{3}}{41}, \frac{20}{41}\sqrt{\frac{2}{3}}\right]_{ortho} = -\frac{5}{41}[1,2,2]_{hex}$.

The shear matrix associated to this vector is given by the formula ShearingMatrix[θ, **v** , **n**] with θ the angle of shear amplitude along the direction of the vector **v**, and normal to the vector **n**. This function can be directly computed with Mathematica. The Rosenbaum's shear matrix is given by the values $\theta = ArcTan(\gamma_2) = 24.26°$, $\mathbf{v} = \mathbf{s_2}$ and , $\mathbf{n} = \left[0, -\frac{c}{a}, \frac{\sqrt{3}}{2}\right]_{ortho}$; it is

$$ShearR = \begin{pmatrix} 1 & 0 & 0 \\ 0 & \frac{1681 - 20\sqrt{246}}{1681} & \frac{15\sqrt{\frac{3}{41}}}{41} \\ 0 & -\frac{160}{41\sqrt{123}} & \frac{1681 + 20\sqrt{246}}{1681} \end{pmatrix} \approx \begin{pmatrix} 1 & 0 & 0 \\ 0 & 0.8134 & 0.09897 \\ 0 & -0.3518 & 1.186 \end{pmatrix}$$

# Figures

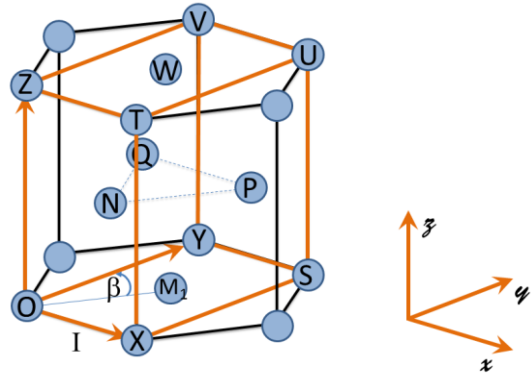

Fig. 1. Hexagonal lattice with its associated orthonormal basis $\mathbf{B}_{ortho} = (\mathbf{x}, \mathbf{y}, \mathbf{z})$, $\mathbf{x} = [100]_{hex}$, $\mathbf{y} = [1\bar{1}0]_{hex}$, $\mathbf{z} = [001]_{hex}$. Some Mg atoms are labelled in order to describe the atomic displacement during the extension twinning.

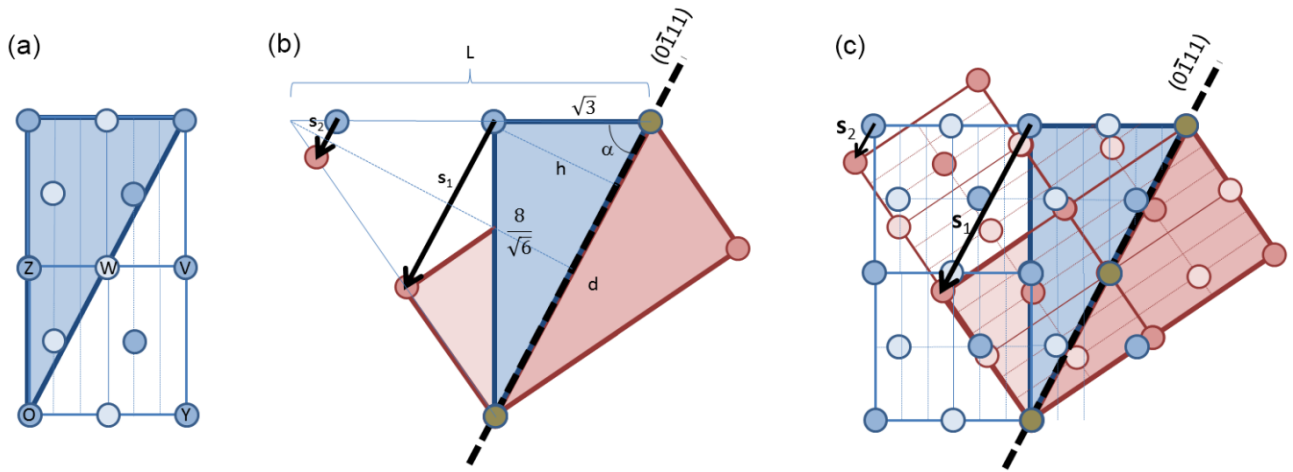

Fig. 2. Shear vectors associated to $(0\bar{1}11)$ twinning. (a) Two unit cells repeated along $z$ and viewed in projection perpendicularly to the $\mathbf{a}$-axis. The $(0\bar{1}11)$ plane appears in projection as the diagonal of the blue triangle. (b) The parent cell (in blue) becomes the twin cell (in salmon) by mirror symmetry through the $(0\bar{1}11)$ plane. The shear vectors $\mathbf{s}_1$ and $\mathbf{s}_2$ reported in literature are indicated by the bold arrows. (c) Same scheme with the atomic positions in the parent crystal and its twin.

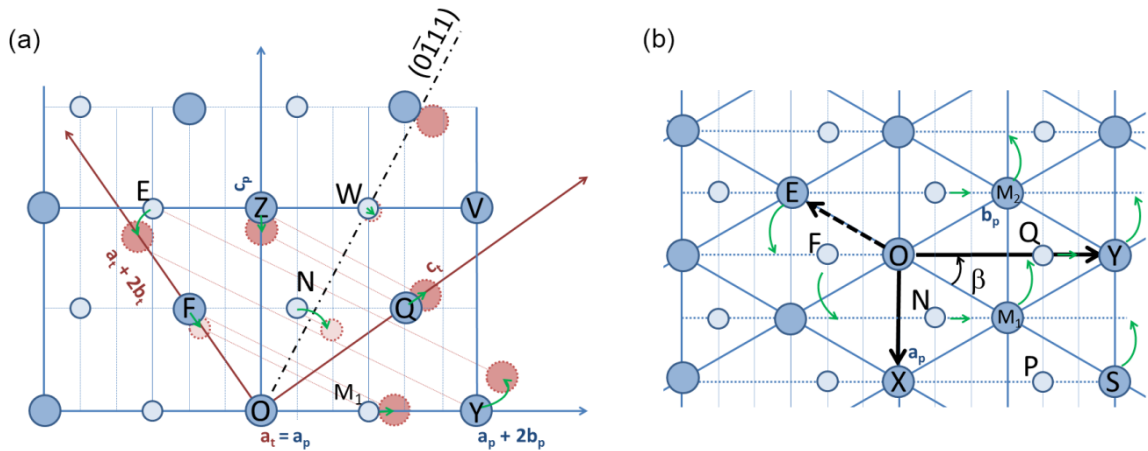

*Fig. 3. Proposition of atom displacements for which the final configuration is hcp and very close to the exact mirror symmetry through the $(0\bar{1}11)$ plane. (a) View on the $(2\bar{1}\bar{1}0)$ plane. The atoms of the parent crystal are in blue and those of the twin in salmon. The dark / light blue and dark / light salmon allows the distinction of atoms on different levels of $(2\bar{1}\bar{1}0)$ planes. The atoms $M_1$, E and F change of level during their movements. The final positions (in salmon) are very close to positions that would be obtained by mirror symmetry the $(0\bar{1}11)$ plane as indicated by the orange lines perpendicular to the $(0\bar{1}11)$ plane. b) Displacements viewed on the (0001) plane. The dark / light blue alloys the distinction of atoms on different levels of (0001) planes.*

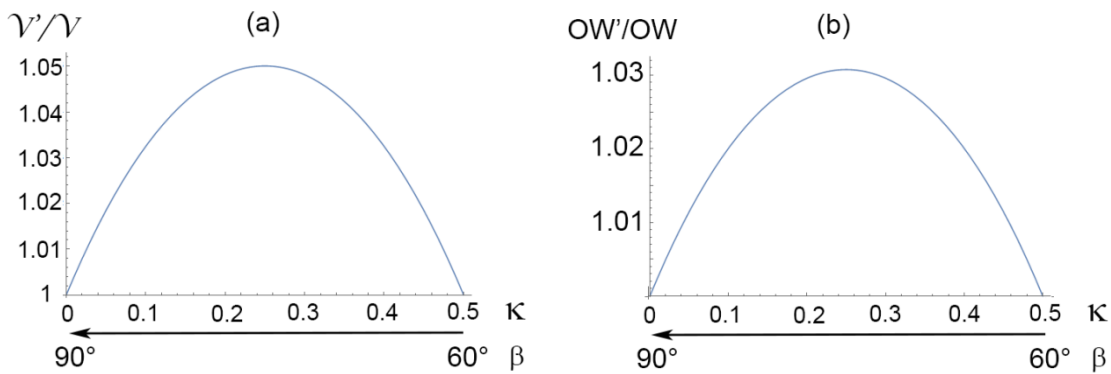

*Fig. 4. Change during contraction twinning of a) volume ratio $\mathcal{V}'/\mathcal{V}$ and b) distances OW'/OW, both function of the parameter $\kappa = Cos(\beta)$, varying from $\kappa_s = 1/2$ i.e $\beta = 60°$ (start) to $\kappa_f = 0$, i.e. e $\beta = 90°$ (finish).*

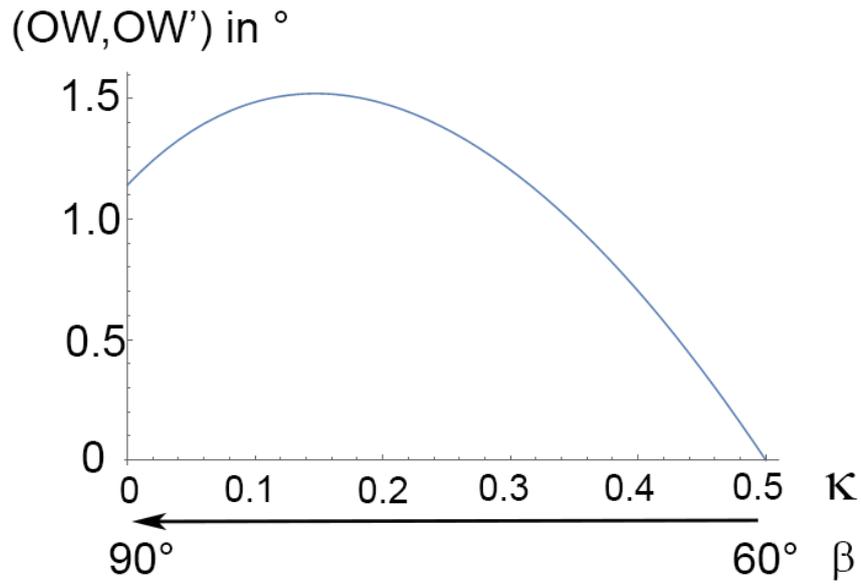

Fig. 5. Evolution of the angle (OW,OW') during twinning as function of the parameter $\kappa = Cos(\beta)$, varying from $\kappa_s = 1/2$ to $\kappa_f = 0$.

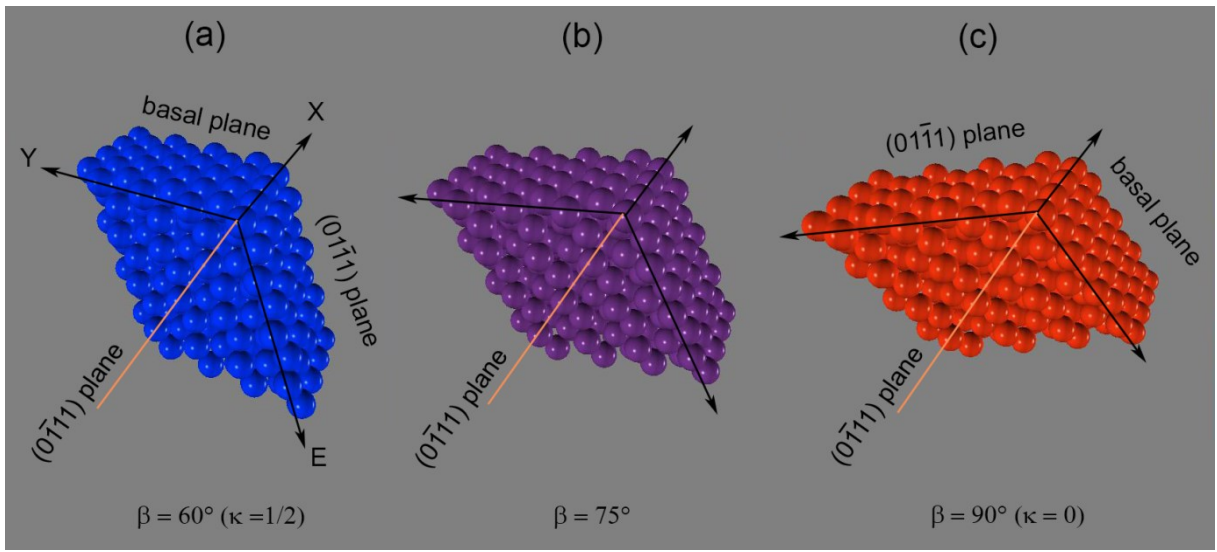

Fig. 6. 3D view of the stretch distortion of a crystal made with 4x4x4 XYE cells. a) Initial hcp structure ($\beta$ = 60°), b) intermediate state ($\beta$ = 75°), and c) final (hcp) structure ($\beta$ = 90°). The corresponding movie is given in Supplementary Materials.

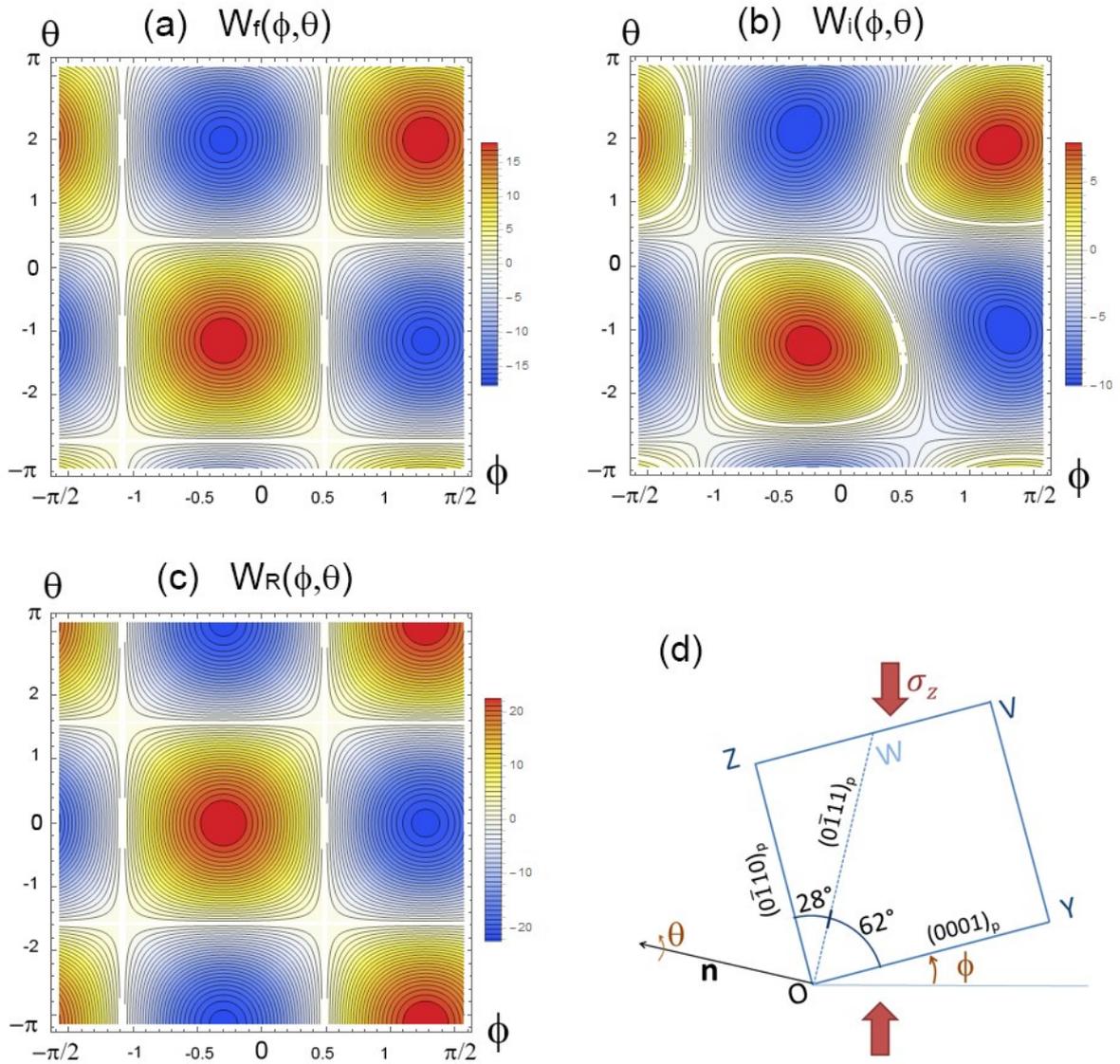

Fig. 7. Interaction work (in MPa) during contraction twinning in a compression stress field of -100 MPa oriented along the **z**-axis. The parent crystal is tilted by an angle $\phi$ around the **a**-axis and rotated by an angle $\theta$ around the **n**-axis (normal to the twinning plane). a) Interaction work $W_f$ calculated with the complete distortion (shear) matrix. $W_f$ is proportional to the a Schmid factor (see Ref. [16]). b) Interaction work $W_i$ calculated with the intermediate distortion matrix corresponding to the maximum volume change. c) Interaction work $W_R$ calculated with the Rosenbaum's shear matrix. d) Schematic view of the orientation of the parent crystal in the external stress field.